\documentclass[showpacs,preprintnumbers,amsmath,nofootinbib,amssymb,superscriptaddress]{revtex4}

\usepackage{amsmath,color}
\usepackage{amssymb}
\usepackage{bm}
\usepackage{graphicx}
\usepackage{multirow}

\usepackage{textcomp}

\allowdisplaybreaks[1]

\newcommand{\nn}{\nonumber \\}
\newcommand{\beq}{\begin{eqnarray}}
\newcommand{\eeq}{\end{eqnarray}}

\begin{document}

\begin{flushright}
YITP-19-46
\end{flushright}

\title{Near threshold $J/\psi$ and $\Upsilon$ photo-production at JLab and RHIC}
\date{}

\author{Yoshitaka Hatta}

\author{Abha Rajan}
\affiliation{Physics Department, Brookhaven National Laboratory, Upton, New York 11973, USA  }

\author{Di-Lun Yang}
\affiliation{Faculty of Science and Technology,  Keio University, Yokohama 223-8522, Japan }
\affiliation{
	Yukawa Institute for Theoretical Physics, Kyoto University, Kyoto 606-8502, Japan\\
}

\date{\today}

\begin{abstract}

We update our previous calculation of $J/\psi$ photo-production near threshold \cite{Hatta:2018ina} by incorporating the recent developments in theory and the new experimental data from the GlueX collaboration at Jefferson Lab \cite{Ali:2019lzf}. 
We then propose to study the near threshold production of $\Upsilon$ and $J/\psi$ in ultraperipheral $pA$ collisions at RHIC.  These processes are sensitive to the gluon condensate in the proton which is  related to the QCD trace anomaly. Our  result emphasizes the role of gluons as the origin of  the proton mass.

\end{abstract}


%
%
%

\maketitle

\section{Introduction}
Although it is a well known fact that the proton has a complex internal structure, much of the confining mechanism that brings the quarks and gluons together to form the proton remains a mystery. In particular, the mass of the proton $M=0.938$ GeV cannot be explained by the naive sum of current quark masses which only accounts for a tiny fraction of the total mass.
The remaining part must come from  the nonperturbative dynamics  of quarks and gluons. Among  various  contributions to the proton mass, the role of the QCD trace anomaly has attracted a lot of attention lately. Dedicated experiments to probe the trace anomaly contribution  are currently running at Jefferson laboratory (JLab) \cite{Joosten:2018gyo}, and similar experiments are planned at the future Electron-Ion Collider (EIC).  

Specifically, JLab measures the photo-production of $J/\psi$ in $ep$ scattering. At low energy, very close to the threshold,  the cross section of this process is sensitive to the gluon condensate $\langle P|F^{\mu\nu}F_{\mu\nu}|P\rangle$ in the proton \cite{Kharzeev:1998bz} which is closely related to the trace anomaly. However, extracting the value of the condensate from the experimental data is highly nontrivial and subject to large systematic uncertainties. This is because  QCD factorization for this process is difficult to establish as it involves the twist-four operator $F^2$, and in practice one has to employ a nonperturbative model to calculate the cross section. Yet, some models allow for a more systematic treatment of the problem than others. In a previous publication \cite{Hatta:2018ina}, two of the present authors have proposed a holographic approach based on gauge/string duality. In the limit of heavy-quark mass, it has been shown that the cross section is directly related to the so-called gravitational form factors of the proton. Since these form factors can be analyzed by other means (e.g., in lattice QCD simulations), a large part of uncertainties associated with the nonperturbative proton matrix elements can be absorbed into those of the form factors. In \cite{Hatta:2018ina}, the theoretical result was fitted to the 40-years-old experimental data from Cornell \cite{Gittelman:1975ix}
and SLAC 
\cite{Camerini:1975cy}  which were the only available data to compare at that time. The quality of the fit was not satisfactory,  especially with the Cornell data which are closer to the threshold. It was not clear whether this was due to the naivety of the model, or perhaps because the old data were not quite accurate. 

Very recently, the GlueX collaboration at JLab has reported new data for the threshold cross section which significantly differ from the Cornell data \cite{Ali:2019lzf}.  Meanwhile, there have been theory developments  on the renormalization of the trace anomaly \cite{Hatta:2018sqd,Tanaka:2018nae} as well as the first lattice calculation of the gluon `D-term'  gravitational form factors \cite{Shanahan:2018pib}. In view of these, we feel it is necessary to revise the calculations and fits in \cite{Hatta:2018ina}. This is what we shall do in the first part of this paper. 

In the second part, we propose a novel way to measure the gluon condensate in experiments. This is the threshold production of $J/\psi$ and $\Upsilon$ in ultraperipheral  $pA$ collisions (UPCs) at RHIC. In UPCs, a heavy nucleus emits almost real photons which interact with the proton electromagnetically. The process thus closely mimics the photo-production limit of $ep$ scattering and serves as nontrivial cross checks of the experimental results as well as the consistency of the theoretical formalism. Moreover, RHIC can study the $\Upsilon$ production which is energetically not possible at JLab. On the other hand, the high energy of RHIC obviously makes the study of threshold production technically difficult. We however argue that this is feasible once the forward upgrade of the STAR detector has been completed \cite{star}.


\section{Nucleon mass decomposition}

The approximate conformal symmetry of the QCD Lagrangian is explicitly broken by the quantum effects. One of the profound consequences of this fact is that the mass of a hadron is directly related to the QCD trace anomaly.  For a single hadron state $|P\rangle$ with mass squared $M^2=P^2$, the QCD energy momentum tensor has the following expectation value 
\beq
\langle P| T^{\alpha\beta}|P\rangle &=& 2P^\alpha P^\beta ,\\
\langle P| T^\alpha_\alpha |P\rangle &=& \langle P| \left( \frac{\beta(g)}{2g}F^{\mu\nu}F_{\mu\nu} + m(1+\gamma_m(g))\bar{\psi}\psi \right)|P\rangle = 2M^2, \label{trace}
\eeq
 where $\beta(g)$ is the beta-function of QCD and $\gamma_m(g)$ is the mass anomalous dimension. The sum over different flavors is implied in $m\bar{\psi}\psi=\sum_f m_f \bar{\psi}_f \psi_f$. It is understood that the vacuum expectation value has been subtracted in the matrix elements. Since hadrons are bound states of quarks and gluons, it is interesting to ask if one can learn more detailed information about the mass structure of hadrons in terms of the quark and gluon degrees of freedom. The partonic decomposition of hadron masses, in particular, the proton mass, has attracted a lot of attention lately both among the theory and experimental communities. On the theory side, the original proposal in \cite{Ji:1994av} was to work in the rest frame of the hadron and decompose, at the operator level,  the time-component of the energy momentum tensor $T^{00}$. This leads to the formula
\beq
M=M_q^{kin}+M_g^{kin}+M_m+M_a. \label{ji}
\eeq   
 where $M_{q/g}^{kin}$ represents the kinetic energy of quarks/gluons, $M_m$ is from the quark mass term, and $M_a$ is the trace anomaly contribution.  
While the decomposition (\ref{ji}) is gauge invariant, the choice of the component $T^{00}$ inevitably brings up the issue of frame dependence. See \cite{Lorce:2017xzd} for a recent attempt to improve on this point. 

In this paper, we propose another  decomposition which is manifestly frame-independent. Instead of decomposing $M$, one can decompose $M^2=P^2$. The trace anomaly in (\ref{trace}) consists of the quark and gluon parts
\beq
T^\alpha_\alpha = (T_q)^\alpha_{\alpha} + (T_g)^\alpha_{\alpha} ,
\eeq
where
\beq
T^{\alpha\beta}_q = i\bar{\psi} \gamma^{(\alpha} D^{\beta)} \psi , \qquad T^{\alpha\beta}_g=-F^{\alpha\lambda}F^\beta_{\ \lambda} +\frac{\eta^{\alpha\beta}}{4}F^2.
\eeq
(The brackets denote symmetrization in indices.) 
 We can thus write 
\beq
M^2= M^2_q + M^2_g, \qquad  M_{q,g}^2=\frac{1}{2} \langle P|(T^R_{q,g} )^\alpha_\alpha |P\rangle. \label{ach}
\eeq
This decomposition makes sense as long as the operators $(T^R_{q,g})^{\alpha}_\alpha$ are carefully defined. They have to be regularized and renormalized in a certain regularization scheme, which means that the  decomposition (\ref{ach}) is scheme dependent. (The sub/super-script $R$ stands for `renormalized.')  While  scheme dependence is always an issue no matter how one decomposes (for example, it is also relevant to (\ref{ji})), the renormalization of $(T^R_{q})^{\alpha}_\alpha$ and $(T^R_{g})^{\alpha}_\alpha$ separately has been investigated only recently in \cite{Hatta:2018sqd,Tanaka:2018nae}, and so far only in dimensional regularization (DR) with the modified minimal subtraction $\overline{\rm MS}$ scheme. 
 Let us briefly recapitulate the main results of \cite{Hatta:2018sqd,Tanaka:2018nae}. In DR, at the bare operator level, the anomaly entirely comes from the gluon part of the energy momentum tensor
\begin{eqnarray}
\langle P|(T_{q})^\alpha_\alpha|P\rangle &=&  \langle P |m\bar{\psi}\psi|P\rangle, \\
\langle P|(T_{g})^\alpha_\alpha|P\rangle &=&  \langle P |\left( m\gamma_m \bar{\psi}\psi + \frac{\beta}{2g} F^2\right)|P\rangle\nonumber \\
&=& (0.637\alpha_s+\cdots) \langle P |m\bar{\psi}\psi|P\rangle +(-0.3583\alpha_s+\cdots) \langle P |F^2|P\rangle, \label{bare}
\end{eqnarray}
where we explicitly show the numerical value of the first term in the perturbative expansion of $\beta(g)/2g$ and $\gamma_m(g)$ for $N_c=3$ and $n_f=3$.  
Under renormalization, the coefficients of this expansion are reshuffled.  This has been worked out to two-loops in  \cite{Hatta:2018sqd} and then extended to three-loops in \cite{Tanaka:2018nae}. Here we quote the result of \cite{Tanaka:2018nae} for $N_c=3$, $n_f=3$ 
	\begin{eqnarray}
		\langle P|(T_{qR})^\alpha_\alpha|P\rangle &=& \mathcal{C}_{qm} \langle P |(m\bar{\psi}\psi)_R|P\rangle 
		+\mathcal{C}_{qF}\langle P |(F^2)_R|P\rangle + {\cal O}(\alpha_s^4),\nonumber \\
		\langle P|(T_{gR})^\alpha_\alpha|P\rangle &=&\mathcal{C}_{gm} \langle P |(m\bar{\psi}\psi)_R|P\rangle +\mathcal{C}_{gF}\langle P |(F^2)_R|P\rangle + {\cal O}(\alpha_s^4),  \label{3}
	\end{eqnarray}
	where 
	\begin{eqnarray}\nonumber
		&&\mathcal{C}_{qm}=1+ 0.14147\alpha_s -0.00823\alpha_s^2 -0.06435\alpha_s^3,\quad
		\mathcal{C}_{qF}=0.07958\alpha_s +0.05887\alpha_s^2 +0.02160\alpha_s^3,
		\\
		&&\mathcal{C}_{gm}=0.49515\alpha_s +0.77659\alpha_s^2 +0.86549\alpha_s^3,\quad
		\mathcal{C}_{gF}=-0.43768\alpha_s -0.26151\alpha_s^2 -0.18383\alpha_s^3.
	\end{eqnarray}
In this formula, both the operators and the running coupling $\alpha_s$ are defined at some  (perturbative) scale $\mu$.  Note that $m\bar{\psi}\psi = (m\bar{\psi}\psi)_R$ in DR, and the renormalization of the operator $F^2_R$ in this scheme is well understood in the literature \cite{Tarrach:1981bi}. 

Once the non-perturbative matrix elements $\langle P|F^2_R|P\rangle$ and $\langle P|(m\bar{\psi}\psi)_R|P\rangle$ are determined by some means,  Eq.~(\ref{ach}) together with (\ref{3}) achieves a manifestly frame-independent, gauge-invariant decomposition of the hadron mass. In the next sections, we shall discuss methods to experimentally  constrain these matrix elements. As a preliminary, here we show how the matrix element of $F^2_R$ is related to  the nucleon's gravitational form factors 
\beq
\langle P'|(T^R_{q,g})^{\alpha\beta}|P\rangle = \bar{u}(P')\Bigl[ A^R_{q,g}\gamma^{(\alpha}\bar{P}^{\beta)} + B^R_{q,g}\frac{\bar{P}^{(\alpha}i\sigma^{\beta)\lambda}\Delta_\lambda}{2M} + C^R_{q,g}\frac{\Delta^\alpha\Delta^\beta-g^{\alpha\beta}\Delta^2}{M} + \bar{C}^R_{q,g}M\eta^{\alpha\beta} \Bigr] u(P) , \label{jid}
\eeq
 where $\bar{P}^\mu\equiv \frac{P^\mu +P'^\mu}{2}$ and $\Delta\equiv P'-P$. $D_{q,g}=4C_{q,g}$ is often called the `D-term'.  All the form factors depend on $\Delta^2$, as well as the renormalization scale $\mu$. 
Taking the trace of (\ref{jid}), we find
\beq
\langle P'|(T^R_{q,g})^\alpha_\alpha|P\rangle = \bar{u}(P')\Bigl[ A^R_{q,g}M + \frac{B^R_{q,g}}{4M}\Delta^2 -3C^R_{q,g}\frac{\Delta^2}{M} + 4\bar{C}^R_{q,g}M\Bigr] u(P). \label{kao}
\eeq
Eliminating $m\bar{\psi}\psi$ from (\ref{3}) and using (\ref{kao}), one finds 
\beq
\langle P'|F^2_R|P\rangle &=& \bar{u}(P') \Bigl[ (K_g A_g^R + K_q A_q^R)M+\frac{K_g B_g^R+K_q B_q^R}{4M}\Delta^2  \nonumber \\ 
&& \qquad \qquad  -3\frac{\Delta^2}{M}(K_g C_g^R + K_q C_q^R) +4(K_g\bar{C}_g^R + K_q \bar{C}_q^R)M\Bigr] u(P), \label{im}
\eeq
where
\beq
K_g= \frac{1}{{\cal C}_{gF} -\frac{{\cal C}_{gm}}{{\cal C}_{qm}}{\cal C}_{qF}}, \qquad K_q= -\frac{{\cal C}_{gm}}{{\cal C}_{qm}}K_g.
\eeq
(\ref{im}) is a useful formula which relates the nonforward matrix element of the operator $F^2_R$ to the gravitational form factors. The latter (excepting $\bar{C}_{q,g}$) have been calculated in lattice QCD simulations.  

 We also comment on the parameter $b$ introduced in 
 \cite{Ji:1994av} 
\beq
b \equiv \frac{\langle P|m(1+\gamma_m)(\bar{\psi}\psi)_R|P\rangle}{2M^2}, \qquad 1-b=  \frac{\langle P|\frac{\beta}{2g}(F^2)_R|P\rangle}{2M^2}. \label{defb}
\eeq
Physically, $b$ is the fraction of $M^2$ which comes from the current quark masses,  analogous to the pion-nucleon sigma term $\sigma \sim \langle P|m\bar{\psi}\psi|P\rangle$. It is scheme and scale-dependent so that one should more properly write $b\to b^R(\mu)$, though we keep the notation $b$ below for simplicity. 
Taking the forward limit of (\ref{im}) and using $\bar{C}_q=-\bar{C}_g$, we find the relation between $b$ and the $\bar{C}_{q,g}$ form factor at zero momentum transfer
\beq
1-b=\frac{\beta(g)}{2g} \left[ (A_g^R(0) + 4\bar{C}_g^R(0)) (K_g-K_q) + K_q\right]. \label{bb}
\eeq	
A recent $n_f=2+1$ lattice calculation at the physical pion mass has found $\frac{\langle P|m(\bar{\psi}\psi)_R|P\rangle}{2M^2} \approx 0.09$ \cite{Yang:2018nqn}. Since $\gamma_m$ is positive, $b$ is slightly larger than this. A simple estimate gives $b\sim 0.13$.  




\section{ $J/\Psi$ production near threshold at JLab }

In this section, we update our previous calculation \cite{Hatta:2018ina} of threshold $J/\psi$ production from holography. The reason  is threefold. Firstly, we use the precise relation (\ref{im}) between the matrix elements of $F^2$ and the gravitational form factors. In \cite{Hatta:2018ina}, the bare relation (\ref{bare}) has been used. Secondly, a lattice QCD calculation of the $C_{g}$-form factor is now available \cite{Shanahan:2018pib}.  Thirdly, very recently the Glue-X collaboration at Jefferson Lab has reported new experimental data on the $J/\psi$ photo-production cross section near threshold \cite{Ali:2019lzf}.  In \cite{Hatta:2018ina}, we have fitted our result to the 40-years-old experimental data from Cornell and SLAC \cite{Gittelman:1975ix,Camerini:1975cy}. The new JLab data seem to be appreciably different from the Cornell data very close to the threshold.

Let us quickly review the discussion of \cite{Hatta:2018ina}. The process of interest is the exclusive production of $J/\psi$ in $ep \to e'\gamma p \to e'p'J/\psi$ near threshold. The intermediate photon state is nearly on-shell (photo-production) with the threshold energy $E_\gamma \approx 8.2$ GeV in the proton rest frame. Since QCD factorization has not yet been established for this process,  the previous works employed various nonperturbative approaches \cite{Kharzeev:1998bz,Brodsky:2000zc,Frankfurt:2002ka,Gryniuk:2016mpk,Hatta:2018ina}.  In \cite{Hatta:2018ina}, two of the present authors proposed a holographic approach in which the scattering between the photon and the proton is described by the graviton and dilaton exchanges in five-dimensional anti-de Sitter  space $AdS_5$. The dilaton is dual to the operator $F^2$ in gauge theory, so the cross section depends on the nonforward matrix element
\beq
\langle P'|F^2|P\rangle, \label{fix}
\eeq
whose forward limit $P'\to P$ is related to the trace anomaly. But this limit is kinematically forbidden, and one has to perform an extrapolation $t=\Delta^2\to 0$.  Refs.~\cite{Kharzeev:1998bz,Gryniuk:2016mpk} assumed vector dominance for $J/\psi$ and related the nonforward matrix element $\langle \gamma(q)|...| J/\psi(k)\rangle$ to a forward matrix element $\langle J/\psi(k)|...| J/\psi(k)\rangle$. However, the validity of vector dominance is unclear for $J/\psi$ \cite{Frankfurt:1985cv}. Moreover,   the momentum transfer near threshold is rather large: At the threshold, $|\Delta| \sim 1.5$ GeV, and this is comparable to the charm quark mass which is treated as heavy. Instead, we tend to agree with the observation in  \cite{Brodsky:2000zc,Frankfurt:2002ka} that the dependence on $t=\Delta^2$ should be that of `two-gluon' form factors, although the authors of \cite{Brodsky:2000zc,Frankfurt:2002ka} did not articulate what exactly these form factors are. Ref.~\cite{Hatta:2018ina} explicitly showed that these are nothing but the gravitational form factors (\ref{jid}), and used the bare relation between $\langle F^2\rangle$ and $A_g,B_g,C_g,\bar{C}_g$. Here we revise the calculation in \cite{Hatta:2018ina} by using the renormalized formula (\ref{im}). Admittedly, this choice (bare or renormalized) is somewhat arbitrary and cannot be unambiguously settled in the framework of \cite{Hatta:2018ina} which does not rely on QCD factorization. Our choice is pragmatic and mostly driven by the necessity to match the lattice QCD results on form factors which are usually presented in the $\overline{\rm MS}$ scheme. We note however, that the difference between (\ref{im}) and the one used in  \cite{Hatta:2018ina} is numerically not significant.

	The cross section is computed as follows \cite{Hatta:2018ina}.  The scattering amplitude for the reaction $\gamma(q) p(P)\rightarrow p(P')J/\psi(k)$ is given by
	\begin{eqnarray}\label{mtr_element}
		\langle P|\epsilon\cdot J|P'k\rangle=X\bar{u}(P')\big[\Pi^{\mu\nu}\Gamma_{\mu\nu}+Y\Pi^{\mu}_{\ \mu}\Gamma\big]u(P),  \label{x}
	\end{eqnarray}
	where 
	\begin{eqnarray}
		\Pi^{\mu\nu}(q,k) \equiv q^{(\mu} k^{\nu)} \epsilon\cdot \xi + \epsilon^{(\mu} \xi^{\nu)} q\cdot k-q^{(\mu} \xi^{\nu)}  k\cdot \epsilon -k^{(\mu} \epsilon^{\nu)} q\cdot \xi.
	\end{eqnarray}
$q^{\mu}(k^{\mu})$ and $\epsilon^{\mu}(\xi^{\mu})$ correspond to the momentum and polarization for $\gamma(J/\psi)$.
The first term corresponds to the graviton exchange, and the second term is from the dilaton exchange. We shall use the value  $Y=-11/80$ from the model used in \cite{Hatta:2018ina}. Explicitly, (c.f., (\ref{im}))
	\begin{eqnarray}
		\Gamma^{\mu\nu}&=&  (A^R_{g}+B^R_{g})\gamma^{(\mu}\bar{P}^{\nu)} -\frac{\bar{P}^\mu \bar{P}^\nu}{M}B^R_{g}   +\frac{1}{3} \left(\frac{\Delta^\mu\Delta^\nu}{\Delta^2}-\eta^{\mu\nu}\right)\left(A^R_g M +\frac{\Delta^2}{4M}B^R_g\right), \label{ga1} \\
		\Gamma&=&\frac{1}{4}\Big[(K_gA^R_{g}+K_qA^R_{q})M+\frac{K_gB^R_{g}+K_qB^R_{q}}{4M}\Delta^2-3\frac{\Delta^2}{M}(K_gC^R_{g}+K_qC^R_{q}) + 4(K_g\bar{C}^R_{g}+K_q\bar{C}^R_{q})M \Big].
	\end{eqnarray}
The differential cross section is given by
	\begin{eqnarray}\label{diff_sigma}
		\frac{d\sigma}{dt}=\frac{\alpha_{\text{EM}}}{4(W^2-M^2)^2}\frac{1}{2}\sum_{\text{pol}}\frac{1}{2}\sum_{\text{spin}}|\langle P|\epsilon\cdot J|P'k\rangle|^2,
	\end{eqnarray}
	where $W^2=(P+q)^2$ and $\alpha_{EM}=e^2/(4\pi)$.  (\ref{diff_sigma}) is proportional to an overall coefficient $X^2$ which is the only fitting parameter in our model. 

We thus use the formula (\ref{im}) with the following recent lattice QCD results for $A^R_g(t,\mu)$ and $C^R_g(t,\mu)$ with $n_f=2+1$ at $\mu=2$ GeV \cite{Shanahan:2018pib}
\beq
A^R_g(t,\mu ) = \frac{0.58}{(1-t/m_A^2)^2},   \qquad C^R_g(t,\mu)=  -\frac{10}{4(1-t/m_C^2)^2},
\eeq
with $m_A=1.13$ GeV and $m_C=0.48$ GeV. 
As in \cite{Hatta:2018ina}, we set $B_g$ to be zero because this form factor is known to be very small numerically. As for $\bar{C}_g^R$, we use the formula  (\ref{bb}) and present the cross section as a function of the unknown parameter $b$ defined in (\ref{defb}). The value of the running coupling is 
\beq
\alpha_s(\mu=2\, {\rm GeV}) = 0.30187,
\eeq
evaluated in the same scheme as in \cite{Tanaka:2018nae}.

\begin{figure}
    \includegraphics[width=0.49\linewidth]{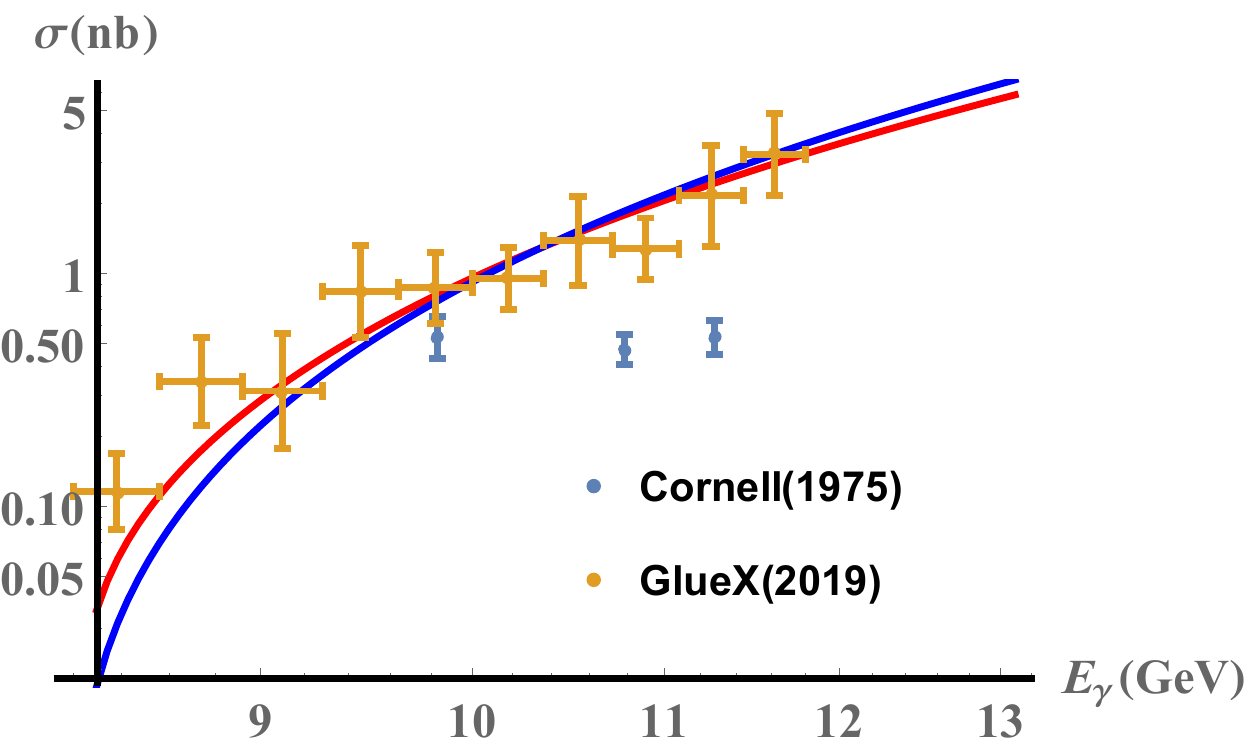}
    \includegraphics[width=0.49\linewidth]{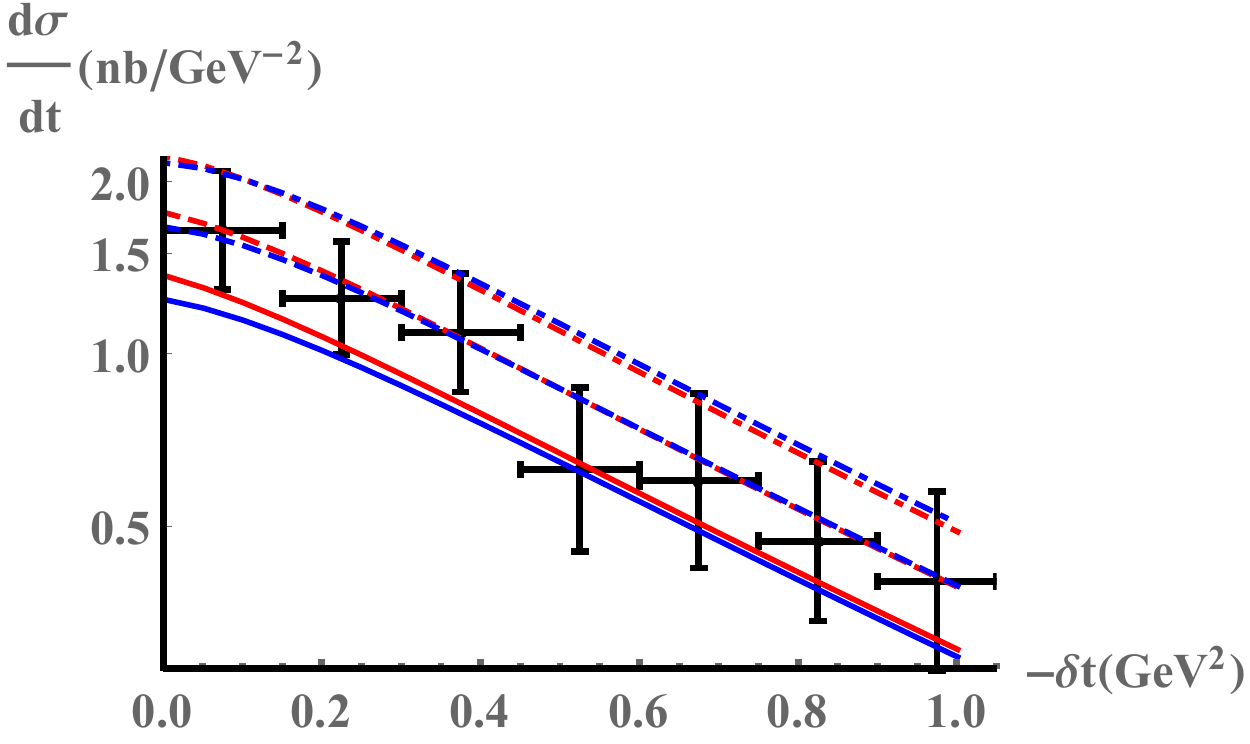}
    \caption{Left panel : Fits of the GlueX data \cite{Ali:2019lzf} for the total cross section. The red curve corresponds to $b=0$ and the blue curve corresponds to $b=1$. Right panel : Comparisons between the GlueX data \cite{Ali:2019lzf} and our model for differential cross sections, where $\delta t=t-t_{\min}$. Color assignments are the same as the left panel, while the solid, dashed, and dot-dashed curves correspond to $E_{\gamma}=10$, $10.3$, and $10.6$ GeV, respectively. The experimental data are taken for $10<E_{\gamma}<11.8$ GeV.}
   \label{glue}
\end{figure}

In Fig.~\ref{glue}, we compare our result with the latest experimental data from the GlueX collaboration \cite{Ali:2019lzf}. The left panel is the energy dependence of the total cross section $\sigma_{tot}$ where old experimental data points from Cornell \cite{Gittelman:1975ix}  are also included. The right panel is the differential cross section $d\sigma/dt$ averaged over a narrow energy interval $10<E_{\gamma}<11.8$ GeV.  The overall normalization is determined from the fit to $\sigma_{tot}$, and the same normalization is used in $d\sigma/dt$.  
 In \cite{Hatta:2018ina}, it was  not possible to fit the Cornell data, so the authors tried to fit the SLAC data which are slightly at higher energies (i.e., further  away from the threshold). However, the region of applicability of the formalism in \cite{Hatta:2018ina} is really limited to  low energies where the scattering amplitude is dominantly real. It is thus gratifying to see that we can now give a reasonable description of the new JLab data. 
As for $d\sigma/dt$, at $E_{\gamma}=10.3$ GeV  our model lie within the experimental error bars for both $b=0$ and $b=1$. However, the difference between $b=0$ and $b=1$ can be merely distinguished only when $-\delta t\approx 0$. When $E_{\gamma}>10.6$ GeV, our result overshoots the measured cross section. In oder to reach a better agreement (or disagreement) between our theoretical result and the experimental observation, it will be  helpful to reduce the interval of photon energy (say, $10<E_{\gamma}<10.6$ GeV ). More importantly, as already noted in  \cite{Hatta:2018ina},   it is highly desirable to go to even lower energies  towards the threshold $E_{\gamma}=8.2$ GeV, because then the difference between the red and blue curves becomes more pronounced.

\begin{figure}
    \includegraphics[width=0.49\linewidth]{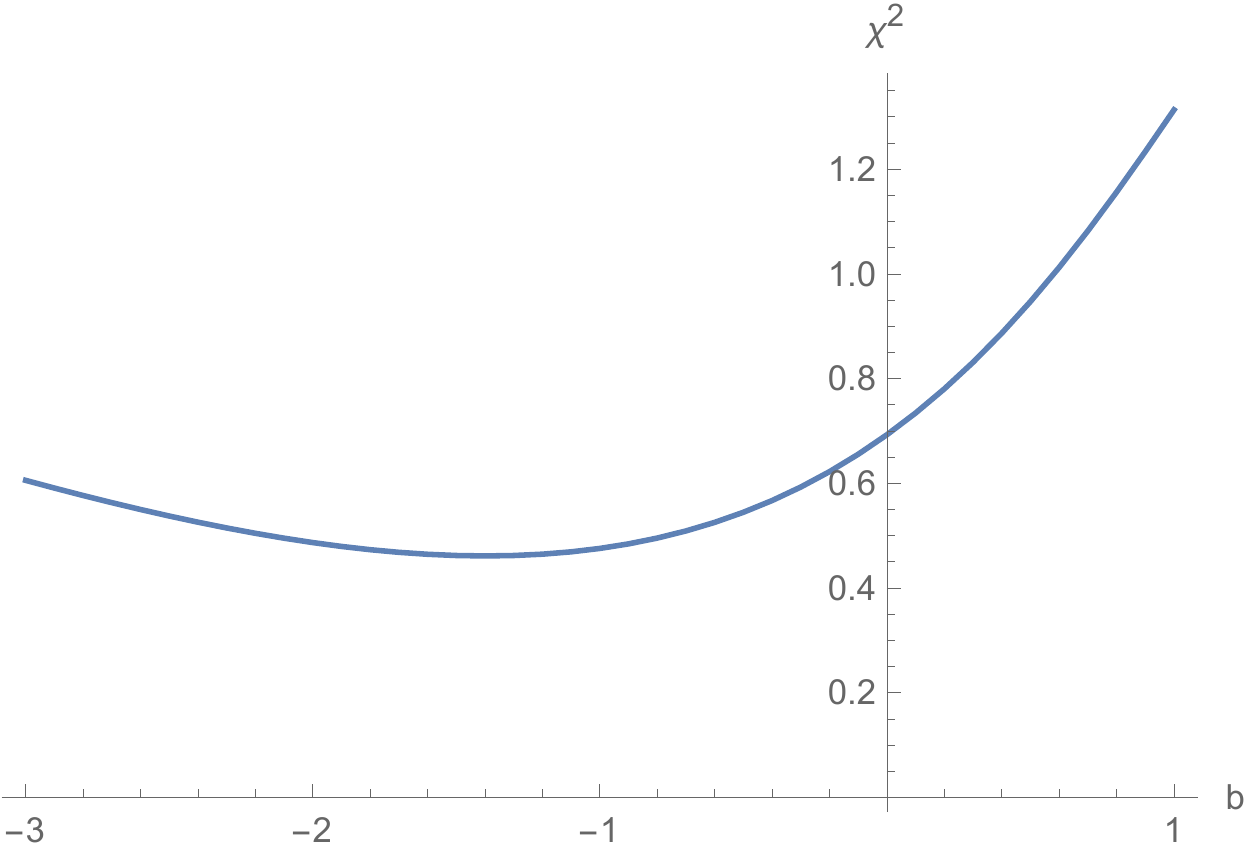}
    \caption{$\chi^2$ as a function of the parameter $b$. }
   \label{chi}
\end{figure}

From the fitting of the total cross section shown in the left panel, one also finds that the $\chi^2$ deviation monotonically decreases when $b$ is reduced within the expected range $0\leq b\leq 1$. Consequently, the maximal anomaly scenario $b=0$ (see (\ref{defb}))  yields the best fit. As a matter of fact, if we allow for negative $b$-values, although this is at odds with the known sign of the nucleon sigma term,   even better fits  can be obtained.  In Fig.~\ref{chi},  we plot $\chi^2$ as a function of $b$.  There is a shallow minimum around $b\sim -1$ and our model is not quite discriminative in the region $b\lesssim 0$.  What we can clearly see, however, is that $\chi^2$ increases steeply towards $b\to 1$,  so  the region $b\sim {\cal O}(1)$ is disfavored. This suggests that the $F^2$ term in the trace anomaly dominates over the quark mass term.   
 

Following (\ref{3}) and (\ref{defb}), we obtain 
	\begin{eqnarray}
		\frac{M_g^2}{M^2}=\Big(\frac{\mathcal{C}_{gm}b}{1+\gamma_m(g)}+\frac{2g\mathcal{C}_{gF}(1-b)}{\beta(g)}\Big)=0.19b+1.23(1-b),
	\end{eqnarray}
with the three-loop formulas for $\gamma_m(g)$ and $\beta(g)$. 
	When $b\leq 0.22$, one finds $(M^R_g)^2\geq M^2$ and thus $(M^R_q)^2\leq 0$. Such a scenario may be foreseen in (\ref{3}) given $\langle P |(F^2)_R|P\rangle <0$. We thus find that, quite interestingly, the quark part of the trace contributes negatively to the nucleon mass (in the present regularization scheme). This further emphasize the role of gluons as the origin of the nucleon mass.


\section{Threshold $J/\psi$ and $\Upsilon$ production in ultraperipheral collisions at RHIC}

In this section, we demonstrate that the threshold production of $J/\psi$ and $\Upsilon$ can be studied also at RHIC. At first sight, this may seem downright impossible since the RHIC energy $\sqrt{s}=200$ GeV is too large to probe any threshold effects. Moreover, RHIC is a collider of protons and heavy nuclei, so superficially it has nothing to do with the physics of photo-production. 

However, it is well known that a heavy nucleus behaves as an abundant source of nearly on-shell photons, called Weisz$\ddot{\rm a}$cker-Williams photons, in  ultraperipheral collisions (UPCs). A UPC is an event in which the impact parameter between the proton and the nucleus is so large that they can interact only via photons emitted from the nucleus.  This process can therefore mimic the photo-production limit of $ep$ scattering.  While the UPC event selection is not as clean as in the case of DIS photo-production, the cross section is enhanced by $Z^2$, the atomic number squared of the nucleus which can be quite large $>{\cal O}(10^3)$. Moreover, at RHIC one can study the threshold production of $\Upsilon$ (the bound state of $b\bar{b}$) which can not be done at JLab because the JLab energy (12 GeV in the proton rest frame) is below the $\Upsilon$ production threshold.   

In UPCs, the cross section $pA \to p'A'J/\psi(k)$ or $pA \to p'A'\Upsilon(k)$  is related to the $p\gamma$ cross section through the standard formula  
\beq
\sigma^{pA}&=&\int d\omega \frac{dN}{d\omega} \sigma^{\gamma p}\nn
&=&\int \frac{d^3k}{2E_k(2\pi)^3} \frac{d^3P'}{2E_{P'}(2\pi)^3}\frac{dN}{d\omega} \frac{e^2}{4MK}  (2\pi)^4 \delta^{(3)}(\vec{P}+\vec{q}-\vec{P'}-\vec{k}) |\langle P|\epsilon \cdot J|P'k\rangle|^2, \label{cross}
\eeq
 where  $\omega=E_{P'}+E_k-E_P$ is the photon energy and 
 \beq
 \frac{dN}{d\omega} = \frac{2Z^2 \alpha_{em}}{\pi \omega}\left[\zeta K_0(\zeta)K_1(\zeta) -\frac{\zeta^2}{2} (K_1^2(\zeta)-K_0^2(\zeta)\right],
 \eeq
is the photon flux.  We defined  $\zeta=\omega \frac{R_p+R_A}{\gamma}$ and  $K=\frac{W^2-M^2}{2M}$, with $W^2=(P+q)^2$ being the $p\gamma$ center of mass energy. 
  We consider $pAu$ collisions at RHIC at $\sqrt{s_{NN}}=200$ GeV and work in the $pp$ center-of-mass frame so that $Z=79$, $\gamma = \sqrt{s_{NN}}/2M\approx 100$ and $R_A \approx 8$ fm. In this frame,
$P^\mu=(E_P,0,0,P)$ ($E_P=\frac{\sqrt{s_{NN}}}{2}$) and $q^\mu=(\omega,0,0,-\omega)$.

The typical value of $\omega$ (from $\zeta\sim 1$) is $\omega\sim 2$ GeV which gives $W=\sqrt{(P+q)^2}\sim 28$ GeV. This is well above the Upsilon production threshold $W\sim 10$ GeV. Due to the asymmetry between the photon and proton energies, the produced quarkonium ($\Upsilon$ or $J/\psi$) with mass $M_Q$ is typically found in the very forward region of the incident proton.  Most of them are produced far away from the threshold. We need to identify the region of phase space corresponding to threshold production and zoom in on that region. 

\begin{figure}
\includegraphics[width=6cm]{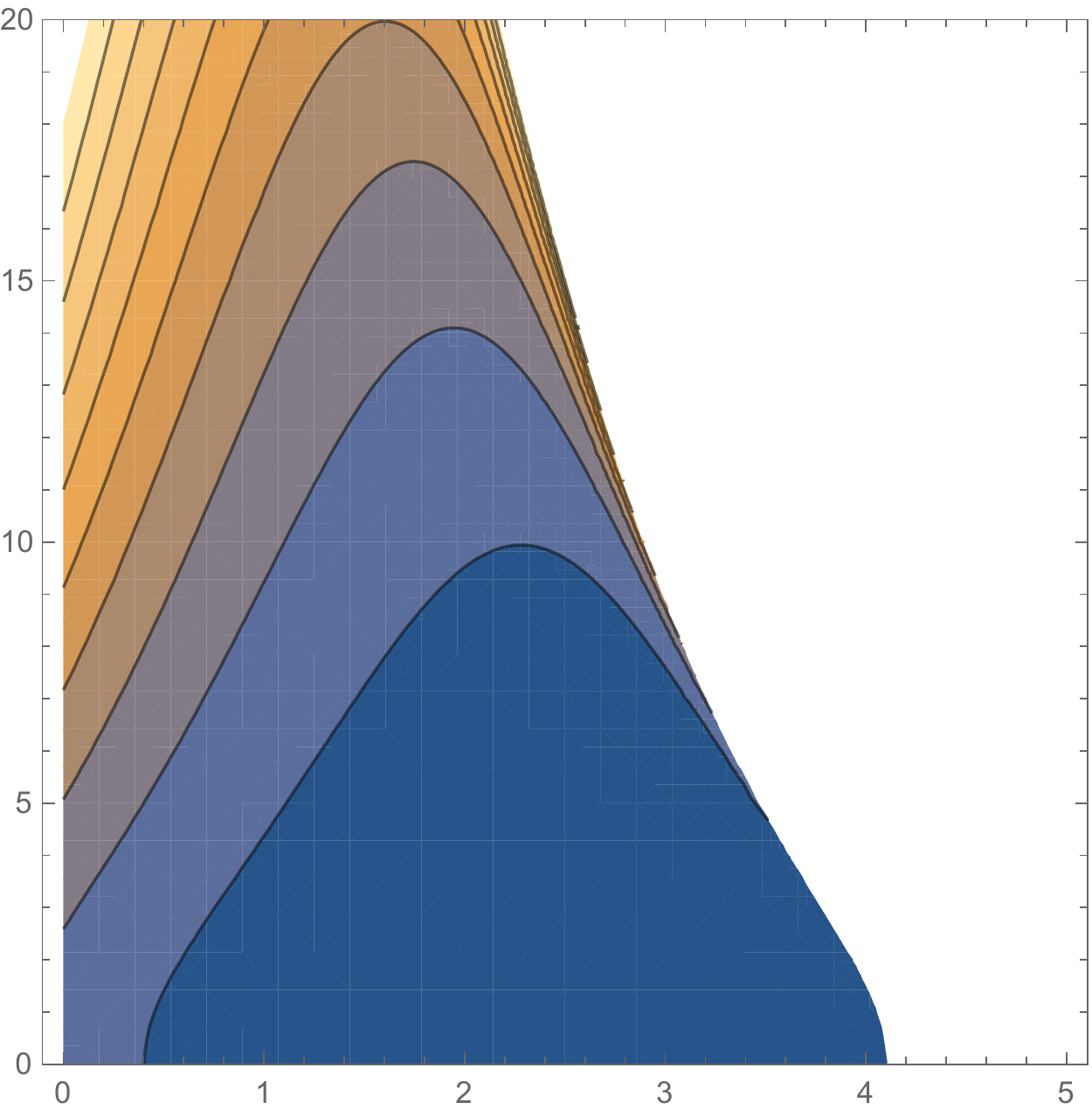}
\space{\qquad\qquad}
\includegraphics[width=6cm]{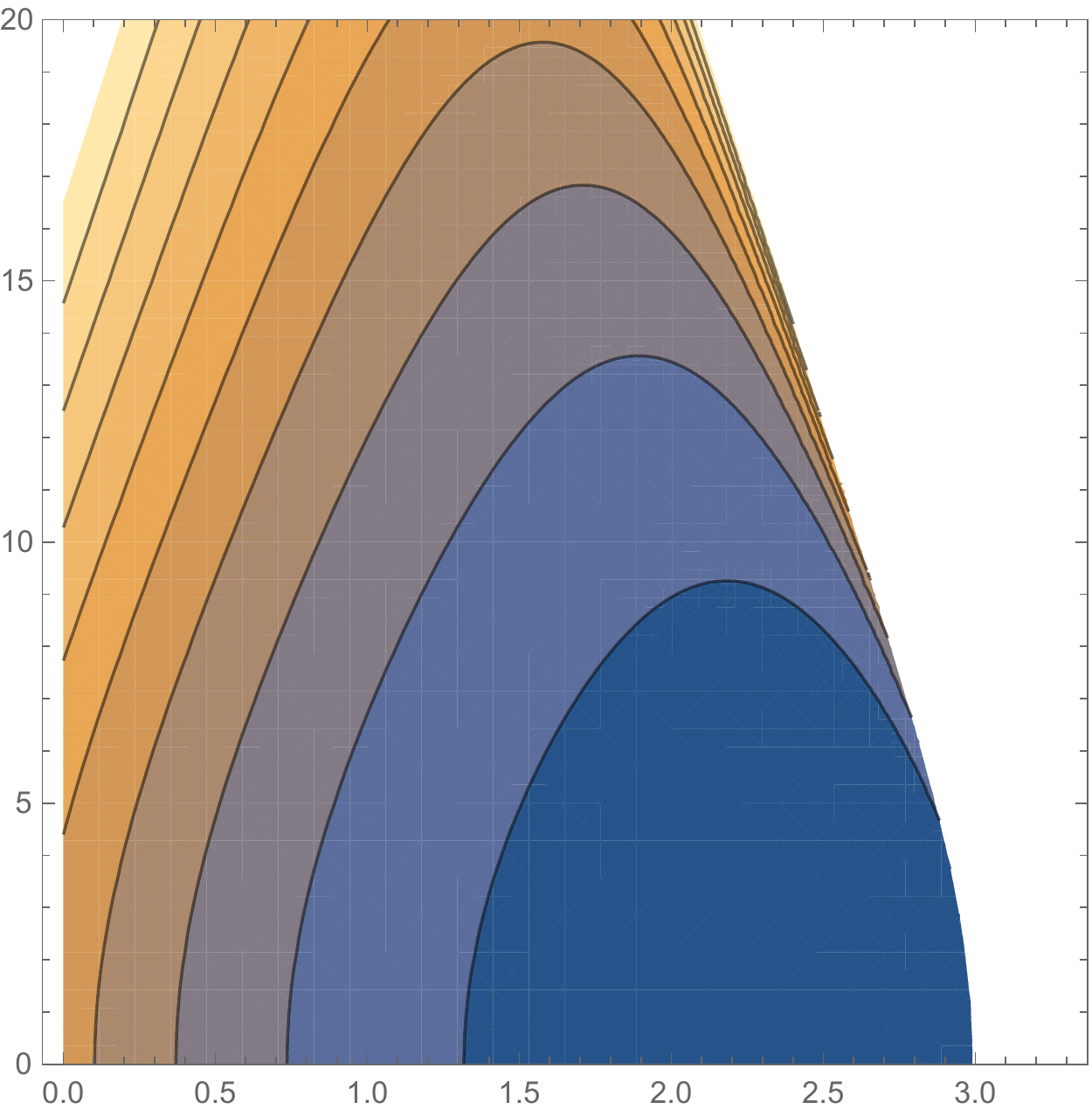} 
\caption{Kinematically allowed region of  $J/\psi$ (left) and $\Upsilon$ (right). The horizontal axis is $y$ and the vertical axis is $k_\perp$. The threshold region is in the lower-right corner. }
\label{contour}
	\end{figure}

Integrating over $\vec{P}'$ in (\ref{cross}) , we get 
\beq
\sigma^{pA}= \frac{ e^2}{64\pi^2M} \int \frac{d^3k}{E_k} \frac{dN}{d\omega} \frac{1}{E_{P'}K} |\langle P|\epsilon \cdot J|P'k\rangle|^2,
\eeq
 where $E_{P'}=\sqrt{M^2+\vec{k}^2}$.  
In terms of the rapidity $y =\frac{1}{2}\ln \frac{E_k+k^3}{E_k-k^3}$ of the quarkonium, we have $d^3k/E_k = dyd^2k_\perp$ so that
 \beq
 \frac{d\sigma^{pA}}{dy d^2k_\perp} =  \frac{ e^2}{64\pi^2M}  \frac{dN}{d\omega} \frac{1}{E_{P'}K} |\langle P|\epsilon \cdot J|P'k\rangle|^2,
 \eeq
 or after averaging over the azimuthal angle, 
  \beq
 \frac{d\sigma^{pA}}{dy dk^2_\perp} =  \frac{\pi e^2}{64\pi^2M}  \frac{dN}{d\omega} \frac{1}{E_{P'}K} |\langle P|\epsilon \cdot J|P'k\rangle|^2.
 \eeq
 In this formula,
 \beq
 E_{P'}&=& \sqrt{M^2+k_\perp^2 +(P-\omega-M^\perp_Q\sinh y)^2 }, \nn
 \omega&=&  \sqrt{M^2+k_\perp^2 +(P-\omega-M^\perp_Q\sinh y)^2  } + M^\perp_{Q}\cosh y -E_P,
 \eeq
 where $M^\perp_Q=\sqrt{k_\perp^2+M_Q^2}$. 
 The second equation can be solved for $\omega$ and the result is 
 \beq
 \omega= \frac{2M^\perp_Q (E_P\cosh y -P\sinh y)-M^2_Q}{2(E_p + P -M^\perp_Q e^y)}.
 \eeq
The threshold condition is 
 \beq
 W^2= M^2+2\omega (E_p+P) > (M+M_{Q})^2 \quad \to \quad  \omega > \frac{M_Q(2M+M_Q)}{2(E_P+P)}  \approx  \begin{cases} 0.27 \, {\rm GeV}  \quad (\Upsilon) \\ 0.039 \, {\rm GeV}\quad  (J/\psi) \end{cases}.
 \eeq
 Therefore, the physical region for the kinematical variables $(y,k_\perp^2)$ is 
\beq
   \frac{2M^\perp_Q (E_P\cosh y -P\sinh y)-M^2_Q}{2(E_p + P -M^\perp_Q e^y)}  - \frac{M_Q(2M+M_Q)}{2(E_P+P)} >0.    \label{fe}
 \eeq
In the left panel of Fig.~\ref{contour}, we  plot the left hand side of (\ref{fe}) for $J/\psi$,  $M_Q=M_\psi=3.10$  with $M=0.94$, $E_P=100$, $P=\sqrt{E_P^2-M^2}$ (all in units of GeV).  The threshold region is the lower-right corner around $y\lesssim 2.9$. The right panel is for $\Upsilon$, $M_Q=M_\Upsilon=9.46$. 

\begin{figure}[ht!]
	\centering
	\begin{tabular}{cc}
		\includegraphics[width=0.48\linewidth]{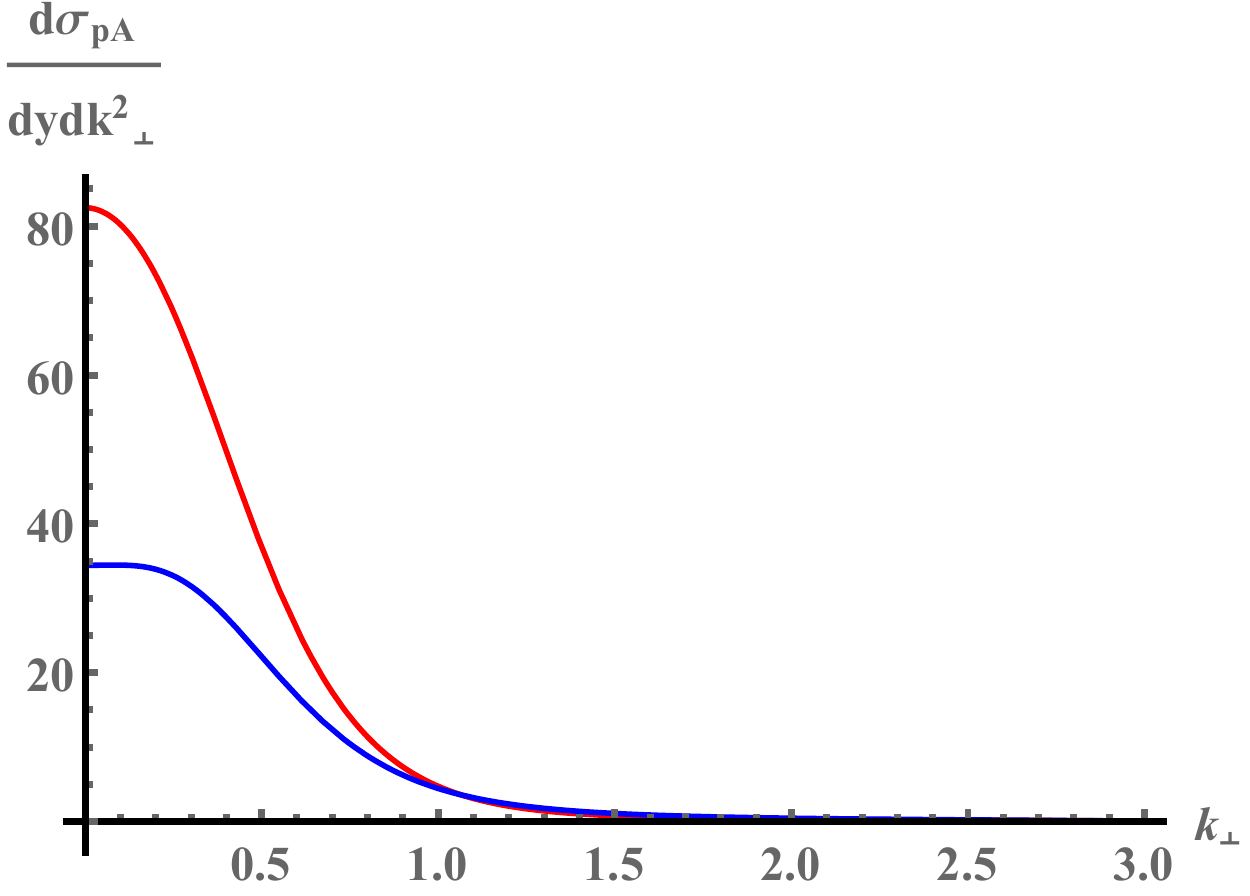} &
		\includegraphics[width=0.48\linewidth]{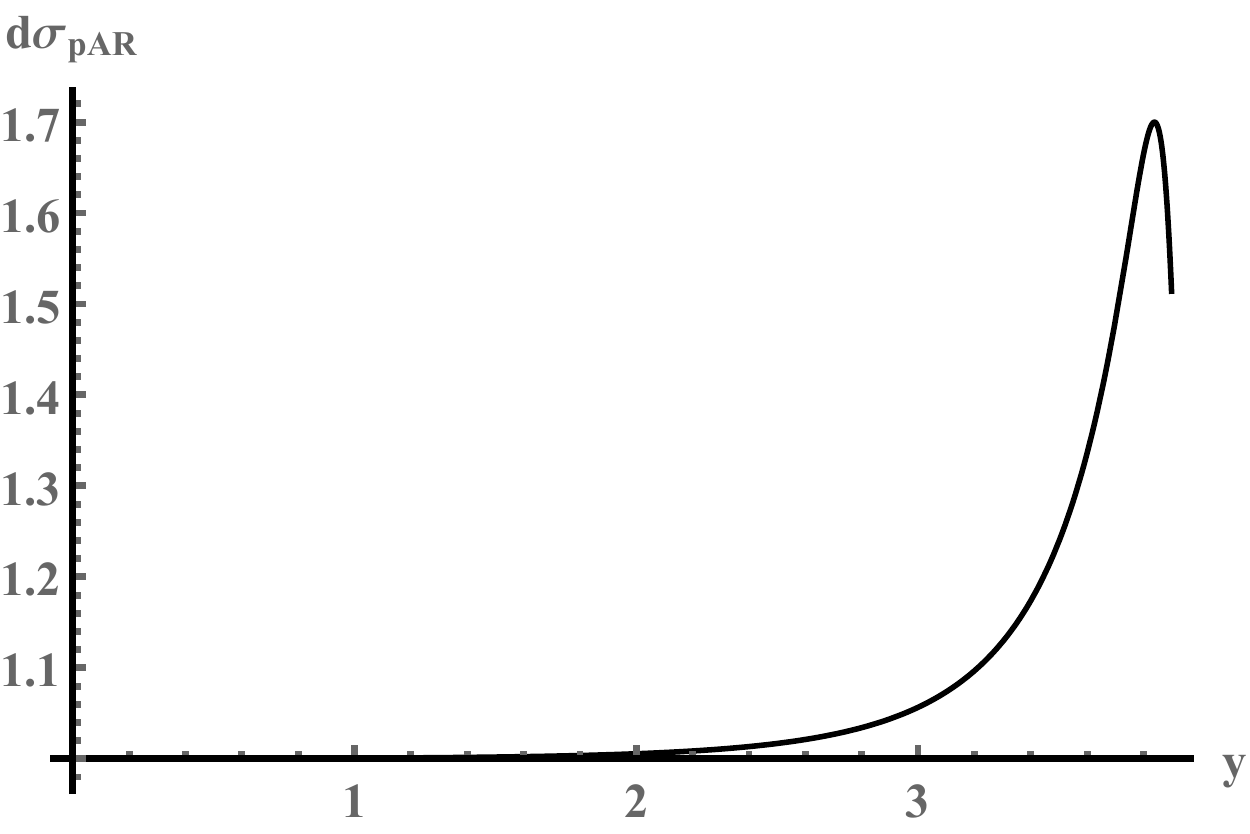}
	\end{tabular}
	\caption{On the left panel, we plot the differential cross sections (nb GeV$^{-2}$) for $J/\psi$ production in UPC at $y=3.8$ with varying $k_{\perp}$ (GeV), where red and blue curves correspond to $b=0$ and $b=1$. On the right panel, we plot the ratio of $d\sigma_{\text{pAR}}\equiv\left(\frac{d\hat{\sigma}_{pA}}{dydk^2_{\perp}}\right)_{b=0}/\left(\frac{d\hat{\sigma}_{pA}}{dydk^2_{\perp}}\right)_{b=1}$ for $J/\psi$ at $k_{\perp}=0.5$ GeV.}
	\label{UPC_dsigma_Jpsi}
\end{figure}

In Fig.~\ref{UPC_dsigma_Jpsi}, left panel, we  show the results of $J/\psi$ production at $y=3.8$. We use the same normalization factor as was used to fit the GlueX data with $b=0$ in the previous section. We immediately notice that the magnitude of the cross section is quite large, thanks to the enhancement factor $Z^2=6241$. The difference between $b=1$ and $b=1$ becomes quite significant as $k_\perp\to 0$. In the right panel, we plot the ratio $d\sigma_{\text{pAR}}\equiv\left(\frac{d\hat{\sigma}_{pA}}{dydk^2_{\perp}}\right)_{b=0}/\left(\frac{d\hat{\sigma}_{pA}}{dydk^2_{\perp}}\right)_{b=1}$ at $k_{\perp}=0.5$ GeV as a function of the rapidity $y$. This plot clearly shows that it is best to focus on the region $y\lesssim 4$. 

\begin{figure}[ht!]
	\centering
	\begin{tabular}{cc}
		\includegraphics[width=0.48\linewidth]{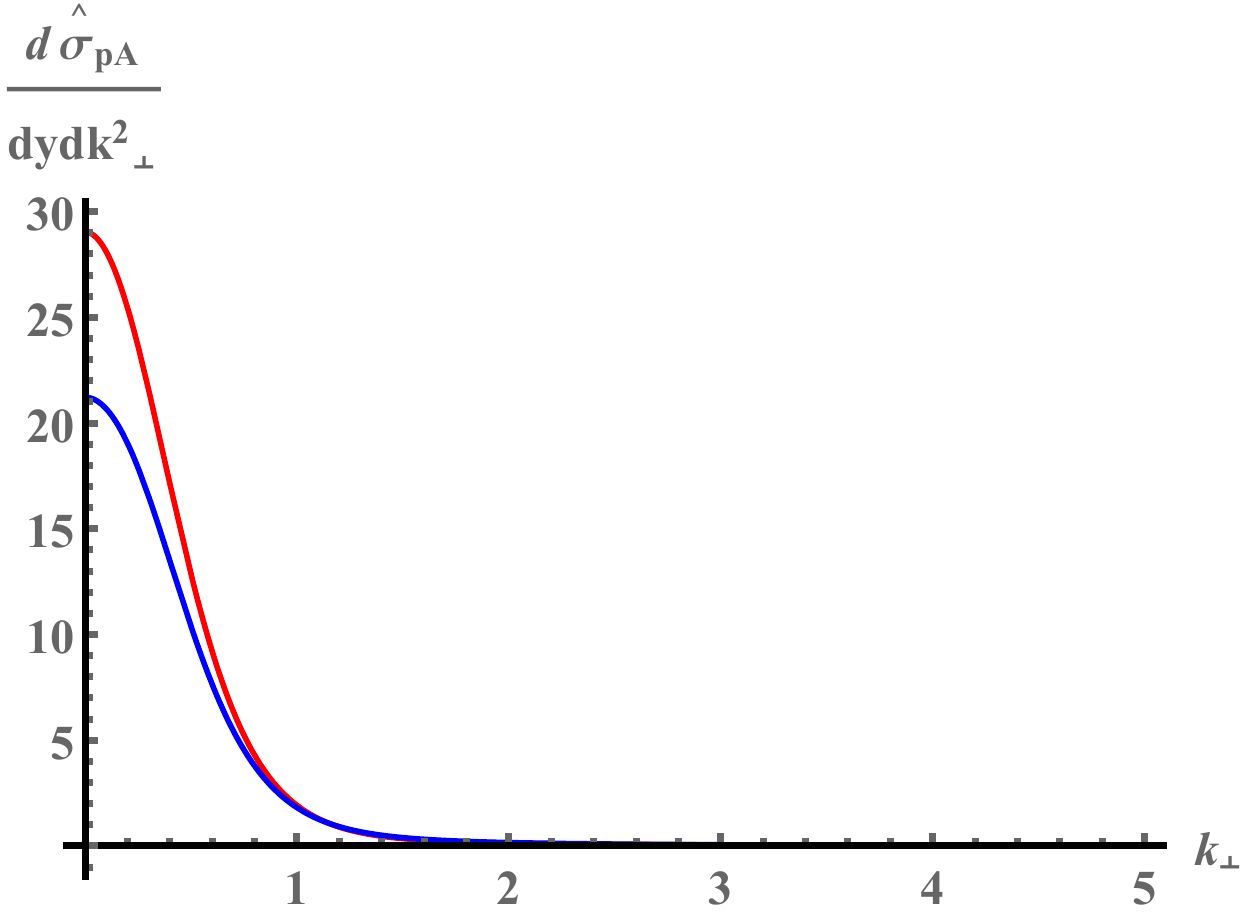} &
		\includegraphics[width=0.48\linewidth]{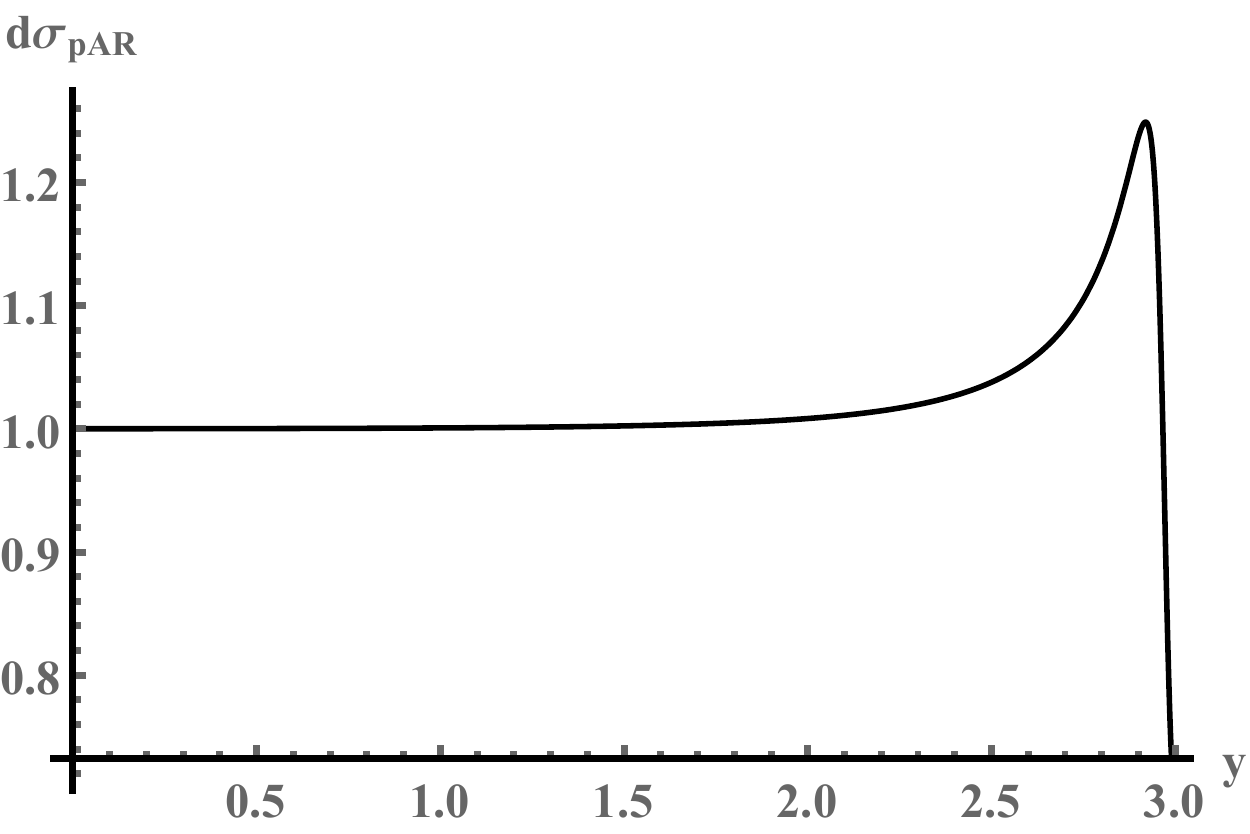}
	\end{tabular}
	\caption{On the left panel, we plot the normalized differential cross sections for $\Upsilon$ production in UPC at $y=2.9$ with varying $k_{\perp}$ (GeV), where red and blue curves correspond to $b=0$ and $b=1$. On the right panel, we plot the ratio of $d\hat\sigma_{\text{pAR}}\equiv\left(\frac{d\hat{\sigma}_{pA}}{dydk^2_{\perp}}\right)_{b=0}/\left(\frac{d\hat{\sigma}_{pA}}{dydk^2_{\perp}}\right)_{b=1}$ at $k_{\perp}=0.5$ GeV.}
	\label{UPC_dsigma_Upsilon}
\end{figure}

In the case of $\Upsilon$, we do not know the normalization factor, as $X$ in (\ref{x}) can depend on the quark mass.\footnote{If one assumes that the cross section scales as $\sigma \sim 1/m_q^2$, one has $\sigma_\Upsilon \sim 0.1\sigma_\psi$, which is not a strong suppression in view of the large $Z^2$ factor.} 
Therefore, we plot the normalized differential cross sections $\hat{\sigma}_{\text{pA}}\equiv\sigma_{\text{pA}}/(\alpha_{EM}X^2)$, on the left panel of Fig.\ref{UPC_dsigma_Upsilon}. In addition, on the right panel, we plot the ratio of $(d\hat{\sigma}_{pA})/(dydk^2_{\perp})$ with $b=0$ to with $b=1$ at fixed $k_{\perp}$. It is found that the near-threshold cross section is enhanced with the maximal anomaly at small $k_{\perp}$ and large rapidity.  Qualitatively, the $J\psi$ production in UPC shares the similar properties as the case for $\Upsilon$.   

We finally note that with the STAR forward upgrade \cite{star} which covers the pseudo-rapidity region $2.5<\eta<4$, the above measurement is feasible. 
Near the threshold, the produced quarkonium typically has high longitudinal momentum $k^3\sim {\cal O}(E_p)$. It can be measured through its decay into a (massless) lepton pair. For a quarkonium with $k_\perp=0$ and $k^3=K$, the produced leptons  have momentum $|k_\perp|=M_Q/2$ and $k^3=K/2$ so that their pseudrapidity $\eta$ is equal to the rapidity $y$ of the parent quarkonium. Fortunately, the relevant values $y\lesssim 2.9$ and $y\lesssim 4$ for $\Upsilon$ and $J/\psi$, respectively, turn out to be perfectly within the coverage of the new detectors.

\section{Conclusions}

In this paper, we first updated our previous fit of the $J/\psi$ photo-production cross section in \cite{Hatta:2018ina}
 in light of the new data from the GlueX collaboration \cite{Ali:2019lzf}. The quality of the fit has improved significantly, and we can now see a hint that the parameter $b$ in (\ref{defb}) is small. This suggests that  the gluon condensate $\sim \langle P|F^2|P\rangle$ dominates over the quark condensate $\langle P|m\bar{\psi}\psi|P\rangle$ in the QCD trace anomaly. In the alternative decomposition (\ref{ach}), it means that the quark part of the trace contributes negatively to the nucleon mass. This observation emphasizes more the role of gluons as the origin of the nucleon mass.  On the other hand, our model is not discriminative enough to determine the value of $b$, and actually, negative values of $b$ are allowed. To fix this problem, it would be very interesting to explore different holographic models from the one considered in \cite{Hatta:2018ina}. 

We then demonstrated that the threshold production can be also studied in ultraperipheral collisions (UPCs) at RHIC in future.  In addition to being complementary to the JLab measurements, a big advantage of RHIC is that one can study the threshold $\Upsilon$ production. The challenge is that one has to measure the quarkonia at very forward rapidities. However, this seems to be doable after  the completion of planned forward upgrade of the STAR detectors.

Finally, it is worthwhile to comment that, although our main target in this paper has been the $\bar{C}_g$ form factor, the near-threshold cross section is very sensitive to the gluon D-term 
\beq
D^R_g(t,\mu) = 4C^R_g(t,\mu),
\eeq
which has attracted considerable interest recently \cite{Burkert:2018bqq,Polyakov:2018zvc} in connection to the `pressure' or `radial force' inside the nucleon. This is because of the explicit prefactor  $\Delta^2$ in (\ref{im}), and $\Delta$ is large near threshold.  For the present purpose, the gluon D-term is an obstruction  to precisely extract the $\bar{C}_g$ contribution. But turning the logic around, it may be possible to use the present processes to constrain the D-term which is poorly known experimentally. We leave this to future works.

\section*{Acknowledgments}
We thank Lubomir Pentchev, Mark Strikman and Kazuhiro Tanaka for discussion and correspondence. This work is supported by the U.~S. Department of Energy, Office of Science, Office of Nuclear Physics under Contracts No. DE-SC0012704.  It is also supported by the LDRD program of  Brookhaven National Laboratory. The
work of D.~Y. is  partially supported by Keio Institute of Pure and Applied Sciences (KiPAS) project in Keio University and Yukawa International Program for Quark-hadron Sciences (YIPQS).

\end{document}